\documentclass[twocolumn,showpacs,showkeys,preprintnumbers,amsmath,amssymb,superscriptaddress]{revtex4}

\usepackage{graphicx}
\usepackage{dcolumn}
\usepackage{bm}

\def\bea{\begin{eqnarray}}
\def\eea{\end{eqnarray}}

\begin{document}

\preprint{Version 5.2}
\keywords{ mean-$p_{t}$ fluctuations, $p_t$ correlations, heavy ion collisions, scale dependence, inverse problems}

\title{Transverse-momentum $p_t$ correlations on $(\eta,\phi)$ from mean-$p_{t}$ fluctuations in Au-Au collisions at $\sqrt{s_{NN}} = $ 200 GeV}

\affiliation{Argonne National Laboratory, Argonne, Illinois 60439}
\affiliation{University of Bern, 3012 Bern, Switzerland}
\affiliation{University of Birmingham, Birmingham, United Kingdom}
\affiliation{Brookhaven National Laboratory, Upton, New York 11973}
\affiliation{California Institute of Technology, Pasadena, California 91125}
\affiliation{University of California, Berkeley, California 94720}
\affiliation{University of California, Davis, California 95616}
\affiliation{University of California, Los Angeles, California 90095}
\affiliation{Carnegie Mellon University, Pittsburgh, Pennsylvania 15213}
\affiliation{Creighton University, Omaha, Nebraska 68178}
\affiliation{Nuclear Physics Institute AS CR, 250 68 \v{R}e\v{z}/Prague, Czech Republic}
\affiliation{Laboratory for High Energy (JINR), Dubna, Russia}
\affiliation{Particle Physics Laboratory (JINR), Dubna, Russia}
\affiliation{University of Frankfurt, Frankfurt, Germany}
\affiliation{Institute of Physics, Bhubaneswar 751005, India}
\affiliation{Indian Institute of Technology, Mumbai, India}
\affiliation{Indiana University, Bloomington, Indiana 47408}
\affiliation{Institut de Recherches Subatomiques, Strasbourg, France}
\affiliation{University of Jammu, Jammu 180001, India}
\affiliation{Kent State University, Kent, Ohio 44242}
\affiliation{Lawrence Berkeley National Laboratory, Berkeley, California 94720}
\affiliation{Massachusetts Institute of Technology, Cambridge, MA 02139-4307}
\affiliation{Max-Planck-Institut f\"ur Physik, Munich, Germany}
\affiliation{Michigan State University, East Lansing, Michigan 48824}
\affiliation{Moscow Engineering Physics Institute, Moscow Russia}
\affiliation{City College of New York, New York City, New York 10031}
\affiliation{NIKHEF and Utrecht University, Amsterdam, The Netherlands}
\affiliation{Ohio State University, Columbus, Ohio 43210}
\affiliation{Panjab University, Chandigarh 160014, India}
\affiliation{Pennsylvania State University, University Park, Pennsylvania 16802}
\affiliation{Institute of High Energy Physics, Protvino, Russia}
\affiliation{Purdue University, West Lafayette, Indiana 47907}
\affiliation{Pusan National University, Pusan, Republic of Korea}
\affiliation{University of Rajasthan, Jaipur 302004, India}
\affiliation{Rice University, Houston, Texas 77251}
\affiliation{Universidade de Sao Paulo, Sao Paulo, Brazil}
\affiliation{University of Science \& Technology of China, Anhui 230027, China}
\affiliation{Shanghai Institute of Applied Physics, Shanghai 201800, China}
\affiliation{SUBATECH, Nantes, France}
\affiliation{Texas A\&M University, College Station, Texas 77843}
\affiliation{University of Texas, Austin, Texas 78712}
\affiliation{Tsinghua University, Beijing 100084, China}
\affiliation{Valparaiso University, Valparaiso, Indiana 46383}
\affiliation{Variable Energy Cyclotron Centre, Kolkata 700064, India}
\affiliation{Warsaw University of Technology, Warsaw, Poland}
\affiliation{University of Washington, Seattle, Washington 98195}
\affiliation{Wayne State University, Detroit, Michigan 48201}
\affiliation{Institute of Particle Physics, CCNU (HZNU), Wuhan 430079, China}
\affiliation{Yale University, New Haven, Connecticut 06520}
\affiliation{University of Zagreb, Zagreb, HR-10002, Croatia}

\author{J.~Adams}\affiliation{University of Birmingham, Birmingham, United Kingdom}
\author{M.M.~Aggarwal}\affiliation{Panjab University, Chandigarh 160014, India}
\author{Z.~Ahammed}\affiliation{Variable Energy Cyclotron Centre, Kolkata 700064, India}
\author{J.~Amonett}\affiliation{Kent State University, Kent, Ohio 44242}
\author{B.D.~Anderson}\affiliation{Kent State University, Kent, Ohio 44242}
\author{D.~Arkhipkin}\affiliation{Particle Physics Laboratory (JINR), Dubna, Russia}
\author{G.S.~Averichev}\affiliation{Laboratory for High Energy (JINR), Dubna, Russia}
\author{S.K.~Badyal}\affiliation{University of Jammu, Jammu 180001, India}
\author{Y.~Bai}\affiliation{NIKHEF and Utrecht University, Amsterdam, The Netherlands}
\author{J.~Balewski}\affiliation{Indiana University, Bloomington, Indiana 47408}
% ...

\collaboration{STAR Collaboration}\noaffiliation

\date{\today}

\begin{abstract}
We present first measurements of the pseudorapidity and azimuth $(\eta,\phi)$ bin-size dependence of event-wise mean transverse momentum $\langle p_{t} \rangle$ fluctuations for Au-Au collisions at $\sqrt{s_{NN}} = 200$ GeV. We invert that dependence to obtain $p_t$ autocorrelations on differences $(\eta_\Delta,\phi_\Delta)$ interpreted to represent velocity/temperature distributions on ($\eta,\phi$). The general form of the autocorrelations suggests that the basic correlation mechanism is parton fragmentation. The autocorrelations vary rapidly with collision centrality, which suggests that fragmentation is strongly modified by a dissipative medium in the more central Au-Au collisions relative to peripheral or p-p collisions.  \\ \\

\end{abstract}

\pacs{24.60.Ky,25.75.Gz}

\maketitle


Central Au-Au collisions at RHIC may generate a color-deconfined medium 
(quark-gluon plasma or QGP)~\cite{QCD}. Some  theoretical descriptions predict abundant low-$p_t$ gluon production in the early stages of high-energy nuclear collisions, with rapid parton thermalization as the source of that medium~\cite{theor0,theor1,theor2}. Particle yields, spectra and high-$p_t$ correlations from Au-Au collisions at $\sqrt{s_{NN}} = $ 130 and 200 GeV provide tantalizing evidence that a colored medium is produced~\cite{qgp,backjet,suppress,suppress2}. Nonstatistical fluctuations of event-wise mean-$p_t$ $\langle p_t \rangle$ \cite{Phenix,ptprl} should help to determine the properties of that medium. A recent measurement of excess $\langle p_t \rangle$ fluctuations in Au-Au collisions at 130 GeV~\cite{ptprl} revealed a large excess of fluctuations compared to independent-particle $p_t$ production. The measurement was obtained at a single {\em scale} (bin size)---the STAR detector acceptance on ($\eta,\phi$) for that analysis. Excess $\langle p_t \rangle$ fluctuations studied with Monte Carlo simulations have been attributed to low-$p_t$ parton fragments (minijets)~\cite{QT}. Measurements of $\langle p_t \rangle$ fluctuations could help to illuminate the role of minijets in nuclear collisions.  

In this paper we report the first measurements of the scale dependence of $\langle p_t \rangle$ fluctuations. Moreover, by inversion of the scale-dependent $\langle p_t \rangle$ variance distribution we obtain $p_t$ {\em autocorrelations}, projections of two-particle distributions on momentum {\em difference variables} $(\eta_\Delta,\phi_\Delta)$, where {\em e.g.,} $\eta_\Delta \equiv \eta_1 - \eta_2$~\cite{inverse}. We compare the resulting $p_t$ correlation patterns to known azimuthal correlations ({\em e.g.,} elliptic flow) and jet angular correlations. We consider the possibility that minijets, as local {\em velocity correlations}, provide a dominating contribution to $p_t$ correlations and quantify centrality dependencies which may describe {\em in-medium modification} of jet correlations. This analysis is based on $\sqrt{s_{NN}} = 200$ GeV Au-Au collisions observed with the STAR detector at the Relativistic Heavy Ion Collider (RHIC).


In each heavy ion collision, and within some region on $(\eta,\phi)$ called a {\em bin}, a number of individual particle $p_t$s is sampled from {\em local} $p_t$ spectra. Local spectrum properties may deviate from the event-averaged $p_t$ spectrum differently at each point on $(\eta,\phi)$ and differently in each event~\cite{mtxmt}. The bin-size (scale) dependence of {\em excess} event-wise $\langle p_t \rangle$ fluctuations measured by {\em variance difference} $\Delta \sigma^2_{p_t:n}(\delta \eta, \delta \phi)$ reflects the correlation structure of the local $p_t$ spectrum properties~\cite{cltps}. Certain aspects of the correlation structure can be accessed when that scale dependence is inverted to obtain $p_t$ autocorrelations~\cite{inverse}: those aspects which depend on relative separation of pairs of points but not on absolute position on $(\eta,\phi)$. The $p_t$ autocorrelations for Au-Au collisions over a range of centralities, their structure and interpretation, are the main subjects of this paper. The next three paragraphs define the $\langle p_t \rangle$ fluctuation measure and outline the derivation of the integral equation which connects its scale variation to the corresponding autocorrelation distribution. Those paragraphs may be omitted in a first reading.

In this analysis the detector acceptance is divided into {\em macro}bins with scales $(\delta \eta,\delta \phi)$. Each macrobin (scales represented by $\delta x$ for brevity) contains in each event some integrated particle multiplicity $n(\delta x)$ and total $p_t(\delta x)$ (scalar sum over particles in the bin). Rather than fluctuations of {\em ratio} $\langle p_t \rangle \equiv p_t / n$ (a source of systematic error) we study fluctuations of {\em difference} $(p_t - n\, \hat p_t) / \sqrt{\bar n}$. The scale-dependent {\em per-particle} $p_t$ variance is defined by $\sigma_{p_t:n}^2(\delta x) \equiv \overline{(p_t(\delta x)-n(\delta x)\, \hat p_t)^2}/\bar n(\delta x)$, where $\hat p_t$ is the inclusive mean particle $p_t$, $\bar n$ is the mean bin multiplicity, \mbox{$p_t:n$} reads $p_t$ {\em given} multiplicity $n$, and the overline represents an average over all macrobins in all events~\cite{ptprl}. The small-scale limit ($\bar n = 1$) of $\sigma_{p_t:n}^2(\delta x)$ is $\sigma^2_{\hat p_t}$, the inclusive single-particle variance. 
The {\em variance difference} is then defined as $\Delta \sigma_{p_t:n}^2(\delta x) \equiv \sigma_{p_t:n}^2(\delta x) - \sigma^2_{\hat p_t}$. 
Variation of $\Delta \sigma^2_{p_t:n}$ on scales $(\delta \eta,\delta \phi)$ corresponds to an integral equation which can be inverted to obtain $p_t$ autocorrelations on difference variables $(\eta_\Delta,\phi_\Delta)$, which compactly represent two-particle $p_t$ correlations on $(\eta,\phi)$~\cite{axialcd} and permit direct interpretation of $\langle p_t \rangle$ fluctuations in terms of physical mechanisms. 

The autocorrelation distribution is a powerful tool for accessing two-particle correlations under certain conditions well satisfied in relativistic nuclear collisions~\cite{axialcd}. An {\em auto}correlation compares a distribution $f(x)$ {\em to itself}. It is effectively a {\em projection by averaging} of product distribution $f(x_1)\cdot f(x_2)$ on $(x_1,x_2)$ onto the difference variable $x_\Delta \equiv x_1 - x_2$. In this analysis we obtain the autocorrelation of the $p_t$ distribution on 2D space $(\eta,\phi)$. Autocorrelations can be determined by {\em pair counting}~\cite{axialcd, axialci}, or by {\em inverting} fluctuation scale dependence to form density ratios following the procedure in~\cite{inverse} first implemented in~\cite{qingjun}. Here we use the latter method. 
 
We can relate variance measurements to autocorrelations in the following way. If a space $x$ is partitioned into {\em micro}bins of fixed size $\epsilon_x$, combined to form macrobins of variable size $\delta x$, the macrobin contents in $\sigma_{p_t:n}^2(\delta x)$ can be expressed as microbin sums, {\em e.g.,} $p_t(\delta x) = \sum_a p_{t,a}(\epsilon_x)$, $a$ being a microbin index. We can then express variance $\sigma_{p_t:n}^2(\delta x)$ as a double sum over microbin indices $(a,b)$ on $(x_1,x_2)$ of terms $\overline{(p_t - n \hat p_t)_a\,(p_t - n \hat p_t)_b}$, which measure the {\em covariance} between bins $a$ and $b$ on $x$ of $p_t$ fluctuations relative to $n \hat p_t$~\cite{mitinv}. As shown in~\cite{inverse,qingjun} we can rearrange the double sum into an outer sum over index $k$ on difference variable $x_\Delta \equiv x_1 - x_2$ ({\em e.g.,} $\eta_1 - \eta_2$, with microbin index $k$) and an inner sum over microbins on sum variable $x_1 + x_2$. The inner sum is $p_t$ difference autocorrelation $\Delta A_k(p_t:n)$ (`difference' referring to $p_t - n\, \hat p_t$). If self pairs are excluded from the microbin sums the $p_t$ difference autocorrelation corresponds to variance {difference} $\Delta \sigma_{p_t:n}^2(\delta x)$. We define reference {\em number} autocorrelation $A_{k,ref}(n)$ as the mean pair number $\bar n^2_k$ in the $k^{th}$ microbin on $x_\Delta$ obtained by averaging products of mean particle numbers  $\bar n_a\, \bar n_b$ along the $k^{th}$ diagonal of $(x_1,x_2)$, that is, with $a-b = k$. That reference is approximately equivalent to the mixed-pair reference autocorrelation which would be obtained by direct pair counting~\cite{inverse}. $A_{k,ref}(n)$ is not obtained explicitly in this analysis, is instead an implicit part of the density ratio defined below and inferred by fluctuation inversion. Autocorrelation {\em densities} $\rho(x_\Delta)$, defined {\em e.g.} by $\Delta A_{k}(p_t:n) \equiv \epsilon^2_x\, \Delta \rho (p_t:n;k\, \epsilon_x)$ and $A_{k,ref}(n) = \epsilon^2_x\,\rho_{ref}(n;k\, \epsilon_x)$, are independent of microbin size. The required per-particle autocorrelation measure corresponding to $\Delta \sigma_{p_t:n}^2(\delta x)$ is density ratio $\Delta \rho(p_t:n) / \sqrt{\rho_{ref}(n)} \equiv \Delta A(p_t:n) / \epsilon_x\, \sqrt{A_{ref}(n)}$ [units (GeV/c)$^2$], which estimates the $p_t$ covariance per particle {\em for a given separation} on $(\eta,\phi)$, averaged over the acceptance~\cite{highptphi}. Within an $O(1)$ constant factor such density ratios have the form of Pearson's {\em correlation coefficient}~\cite{pearson}: the average covariance for all pairs of bins with a {\em given separation} on $(\eta,\phi)$ relative to the geometric mean of Poisson number variances for those bin pairs.


For this 2D scaling analysis we generalize $\delta x \rightarrow  (\delta \eta,\delta \phi)$ to obtain the per-particle conditional $p_t$ variance difference (also defining difference factor $\Delta \sigma_{p_t:n}$~\cite{ptprl}) as the 2D discrete integral equation 
\bea \label{inverse}
\Delta \sigma^2_{p_t:n}(m_\delta \, \epsilon_\eta, n_\delta \, \epsilon_\phi) &\equiv& 2 \sigma_{\hat p_t} \Delta \sigma_{p_t:n}(m_\delta \, \epsilon_\eta, n_\delta \, \epsilon_\phi) \\ \nonumber
= 4 \sum_{k,l=1}^{m_\delta,n_\delta} \epsilon_\eta \epsilon_\phi & &\hspace{-.25in} K_{m_\delta n_\delta;kl}  \,   \frac{\Delta \rho(p_t:n;k\,\epsilon_\eta, l\, \epsilon_\phi) }{ \sqrt{\rho_{ref}(n;k\,\epsilon_\eta, l\, \epsilon_\phi)}} ,
\end{eqnarray}
with kernel $K_{m_\delta n_\delta;kl} \equiv (m_\delta - {k + 1/2})/{m_\delta} \cdot (n_\delta-{l+1/2})/{n_\delta}$. That integral equation can be inverted to obtain autocorrelation density ratio ${\Delta \rho_{}}/{ \sqrt{\rho_{ref}}}$ 
as a per-particle $p_t$ correlation measure on $(\eta_\Delta,\phi_\Delta)$ from the scale dependence of $\langle p_t \rangle$ fluctuations represented by variance difference $\Delta \sigma^2_{p_t:n}(\delta \eta,\delta \phi)$~\cite{inverse,mitinv}.


Data for this analysis were obtained with the STAR detector~\cite{starnim} using a 0.5~T uniform magnetic field parallel to the beam axis. Event triggering  and charged-particle measurements with the time projection chamber (TPC) are described in \cite{starnim}. Track definitions, tracking efficiency and background corrections, event and track quality cuts and primary-particle definition are similar to those described in~\cite{ptprl,spectra} for 130 GeV. While there are quantitative differences between the two energies and detector configurations, the better quality of full-magnetic-field tracking at 200 GeV tends to more than offset the effect of larger track densities there compared to the half-field tracking at 130 GeV. The difference between backgrounds is a few percent of the total track yield (larger for 200 GeV) and is included in the corrections. Tracks were accepted with pseudorapidity in the range $|\eta| < $ 1, transverse momentum in the range $p_t \in [0.15,2]$ GeV/c and $2\pi$ azimuth, defining the detector acceptance for this analysis. Particles identification was not implemented. Eleven centrality classes were defined as fractions of $\sigma_{tot}$ (nine equal fractions from 90\% to 10\%, the top 10\% being further divided in half). The centralities specified below, rounded to the nearest 5\%, are within 2\% of the defined values. Centralities were determined using the uncorrected number $N$ of charged particles in $|\eta|< 1$~\cite{central}.

\begin{figure}[h]
\begin{tabular}{cc}
\begin{minipage}{.47\linewidth}
\includegraphics[keepaspectratio,width=1.65in]{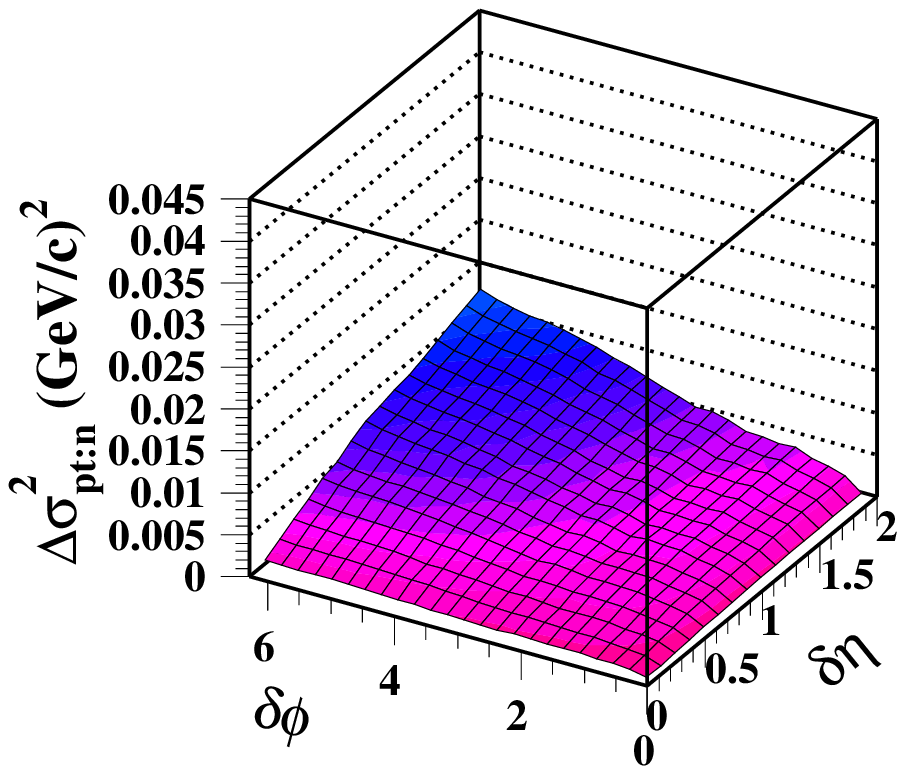}
\end{minipage} &
\begin{minipage}{.47\linewidth}
\includegraphics[keepaspectratio,width=1.65in]{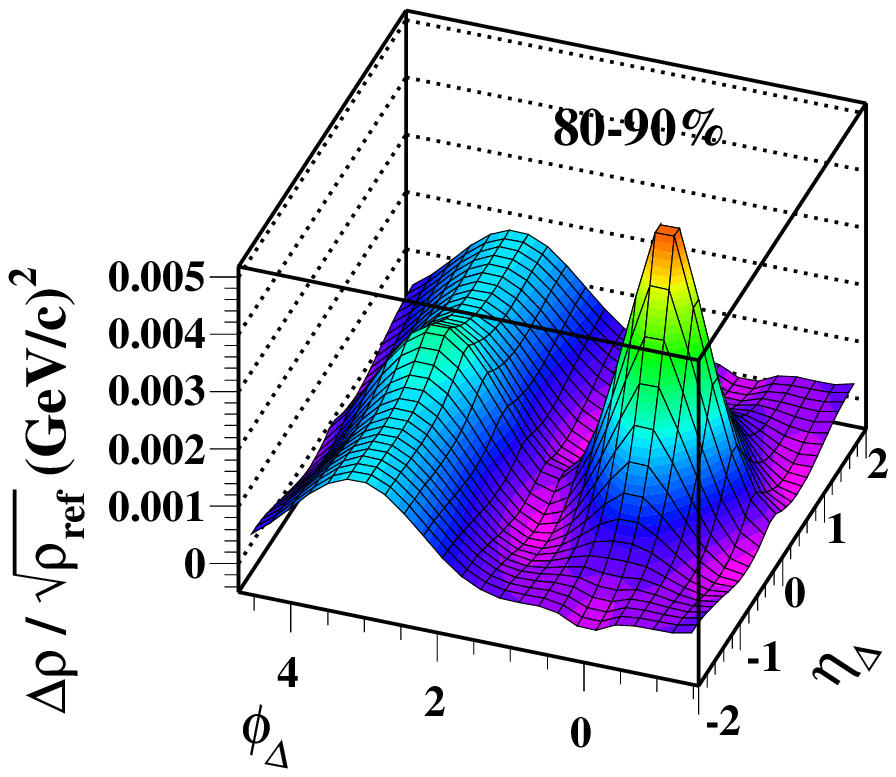}
\end{minipage}\\
\begin{minipage}{.47\linewidth}
\includegraphics[keepaspectratio,width=1.65in]{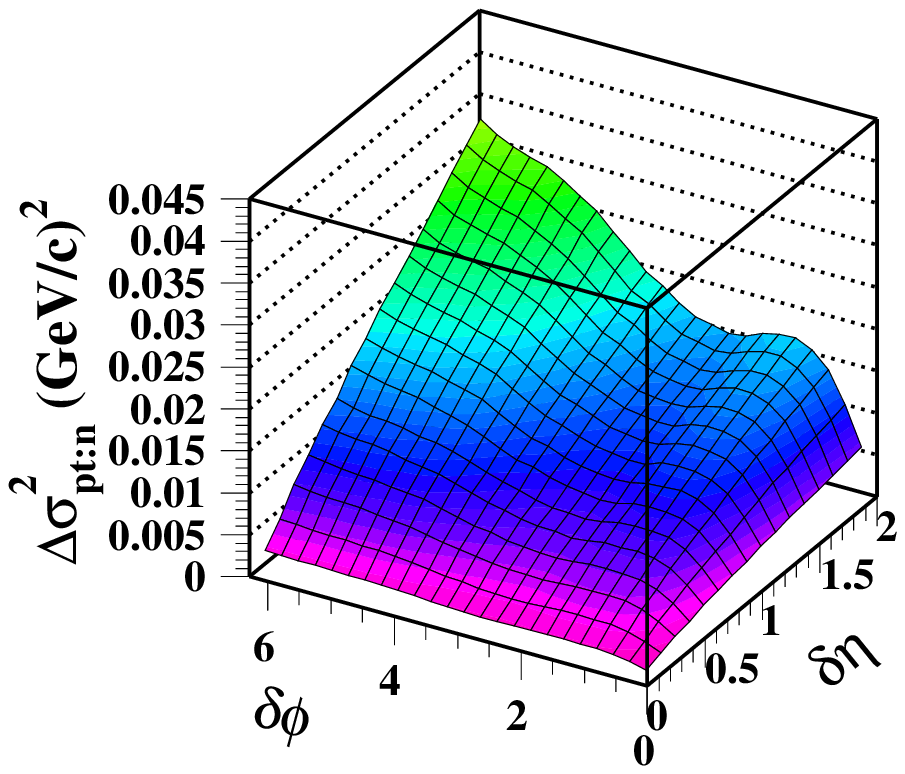}
\end{minipage} &
\begin{minipage}{.47\linewidth}
\includegraphics[keepaspectratio,width=1.65in]{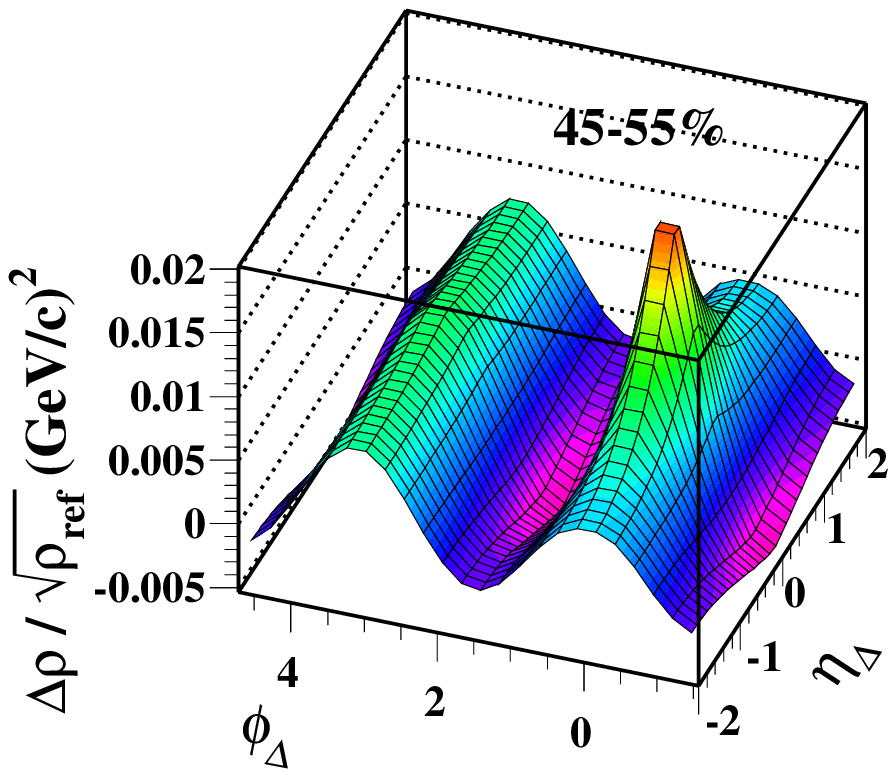}
\end{minipage}\\
\begin{minipage}{.47\linewidth}
\includegraphics[keepaspectratio,width=1.65in]{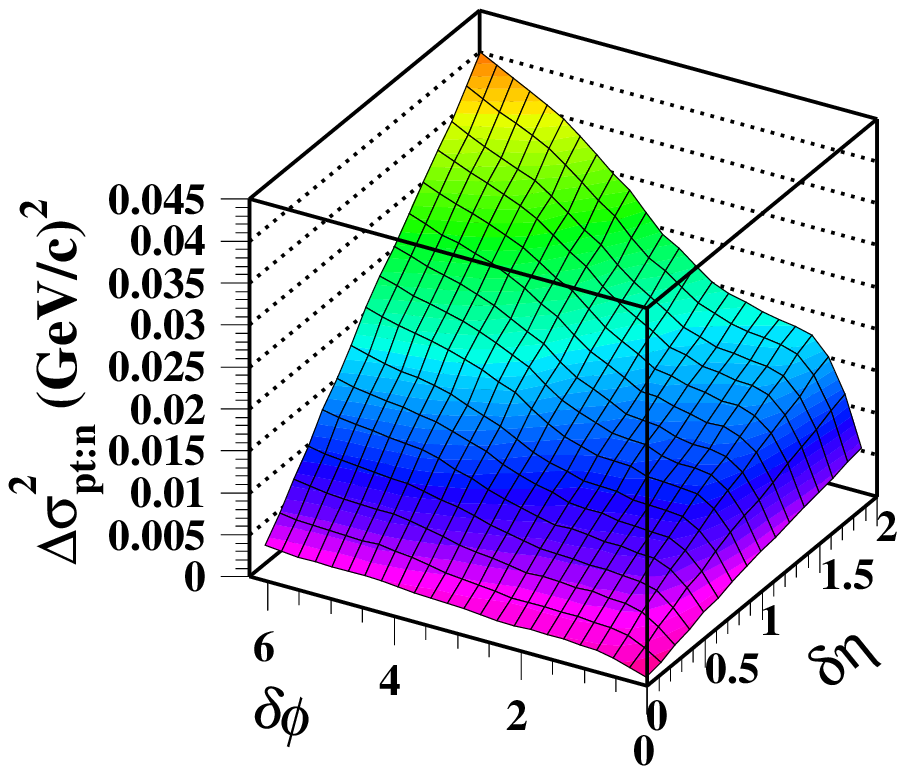}
\end{minipage} &
\begin{minipage}{.47\linewidth}
\includegraphics[keepaspectratio,width=1.65in]{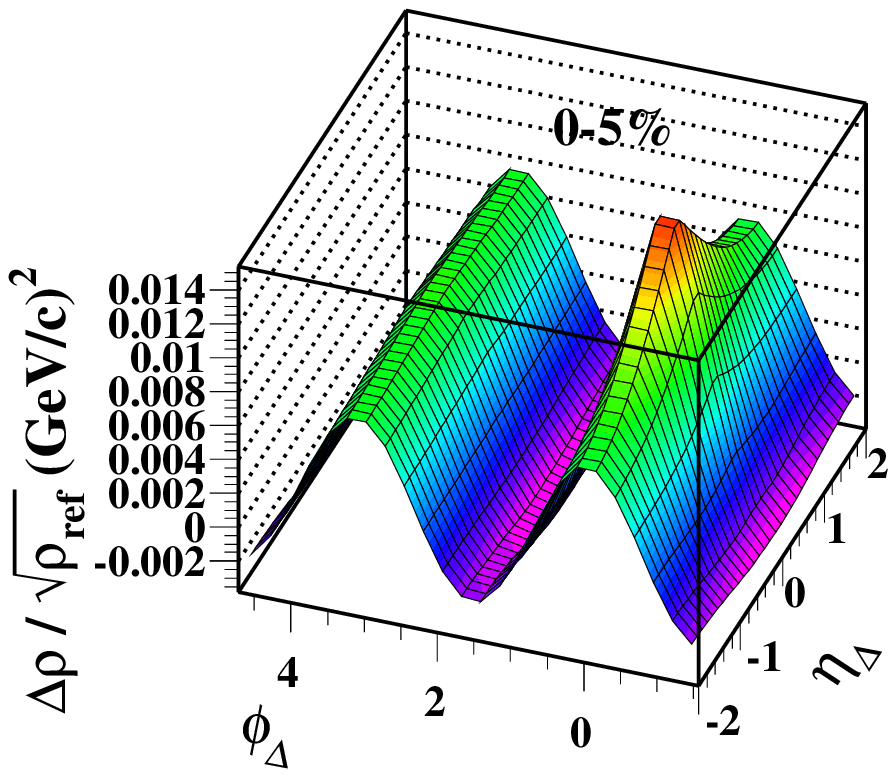}
\end{minipage}\\
\end{tabular}
\caption{Left panels: $\Delta \sigma^2_{p_{t}:n}$ (GeV/c)$^2$ distributions on scale $(\delta \eta,\delta \phi$) for three centrality bins: 80-90\% of total cross section (top panel), 45-55\% of total cross section (middle panel) and 0-5\% of total cross section (bottom panel). Right panels: Corresponding autocorrelations on difference variables ($\eta_{\Delta},\phi_{\Delta}$).\label{fig1}}
\end{figure}

Fig.~\ref{fig1} (left panels) shows the scale dependence of variance difference $\Delta \sigma^2_{p_t:n}(\delta \eta,\delta \phi)$ in Eq.~\ref{inverse} for three of the eleven centralities analyzed. The scale axes are divided into microbins: 16 on pseudorapidity scale $\delta \eta$ and 24 on azimuth scale $\delta \phi$. Variance differences typically increase monotonically with $\delta\eta$ but have more complex behavior on $\delta\phi$. Measurements of difference factor $\Delta \sigma_{p_t:n}$ at $\sqrt{s_{NN}} = 130$ GeV reported in \cite{ptprl} correspond to the single point at the STAR acceptance scale ($2,\, 2\pi$) for each centrality. 
To access the underlying dynamics we extract the corresponding autocorrelation distributions.
Fig.~\ref{fig1} (right panels) shows 2D autocorrelations (by construction symmetric about $\eta_\Delta,\phi_\Delta = 0$) inferred from fluctuation scale dependence in the left panels by inverting Eq.~(\ref{inverse}) \cite{inverse}. Autocorrelations have distinct same-side ($|\phi_\Delta| < \pi/2$) and away-side ($|\phi_\Delta| > \pi/2$) components. For peripheral collisions (top-right panel) the same-side peak appears to be nearly symmetric on $(\eta_\Delta,\phi_\Delta)$, however, {\em cf.} the peak widths in Fig.~\ref{fig3}. In general, the correlation structure evolves rapidly with centrality. 


Errors for $\langle p_t \rangle$ fluctuation measurements in Fig.~\ref{fig1} (left panels) are discussed in \cite{ptprl}. Statistical errors for those variance differences are typically less than 0.001 (GeV/c)$^2$ for all scales and centralities. The inversion process (effectively a differentiation, which acts as a `high-pass' filter) tends to exaggerate small-wavelength noise on the autocorrelation. Control of that noise during inversion requires a standard procedure called {\em regularization}, in which each bin of ${\Delta \rho / \sqrt{\rho_{ref}}}$ is treated as a $\chi^2$ fitting parameter, incorporating a {smoothing} term with corresponding Lagrange multiplier into the $\chi^2$ expression~\cite{inverse,qingjun}. Autocorrelation errors then have two components: statistical noise which survives smoothing and systematic error due to image distortion by smoothing. Statistical errors on the autocorrelation are estimated by inverting the noise estimate for $\Delta \sigma^2_{p_t:n}$. The per-bin {\em r.m.s.} statistical error which survives smoothing is about 0.0002 (GeV/c)$^2$ for all autocorrelations. The smoothing distortion, estimated by passing data through the inversion process twice and comparing the resulting two autocorrelation versions~\cite{inverse,qingjun}, typically peaks at about 5\% of the maximum autocorrelation value at points of maximum gradient. Correlation amplitudes inferred from model fits (see below) were corrected for tracking inefficiencies and background contamination~\cite{ptprl}.  An overall systematic error of $\pm 14$\% for corrected amplitudes reflects uncertainty in extrapolation of variance-difference measurements to the true number of primary particles in the acceptance.

\begin{figure}[h]
\begin{tabular}{cc}
\begin{minipage}{.47\linewidth}
\includegraphics[keepaspectratio,width=1.65in]{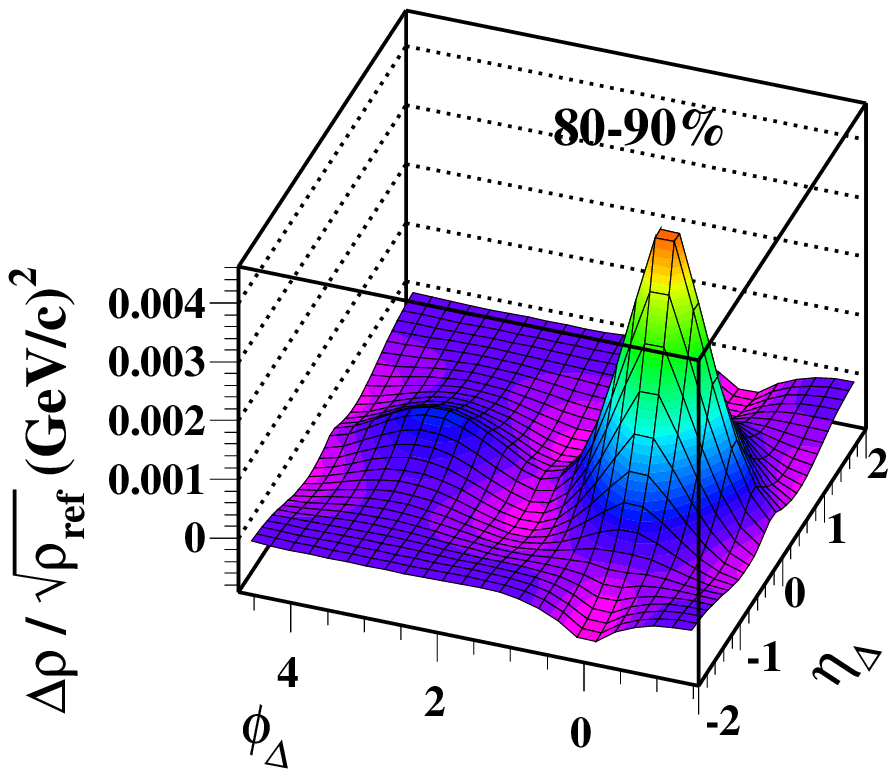}
\end{minipage} & 
\begin{minipage}{.47\linewidth}
\includegraphics[keepaspectratio,width=1.65in]{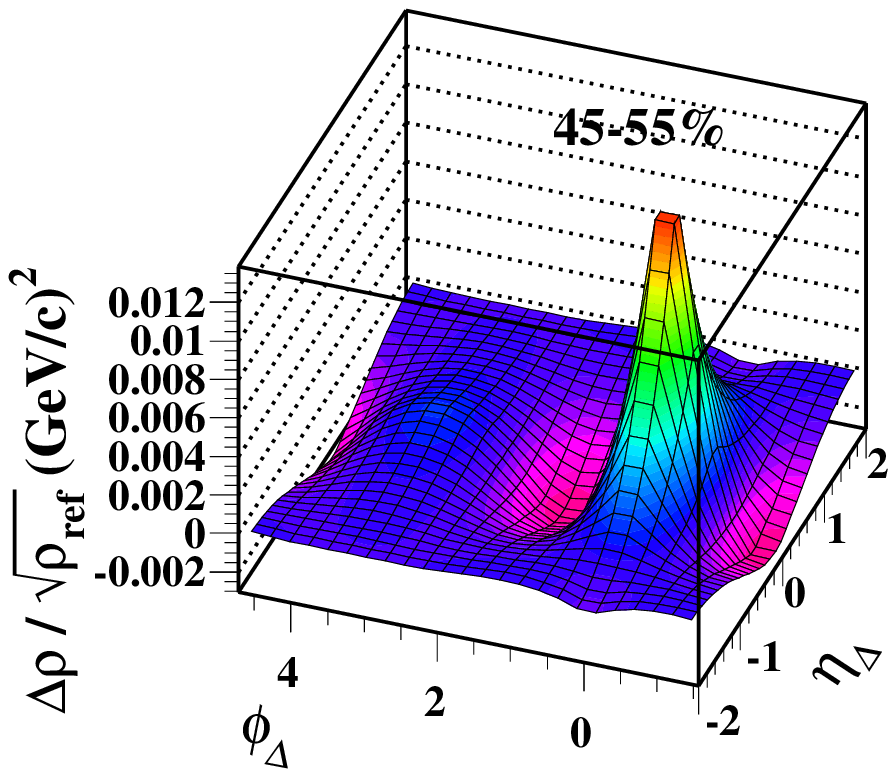}
\end{minipage}  \\ 
\begin{minipage}{.47\linewidth}
\includegraphics[keepaspectratio,width=1.65in]{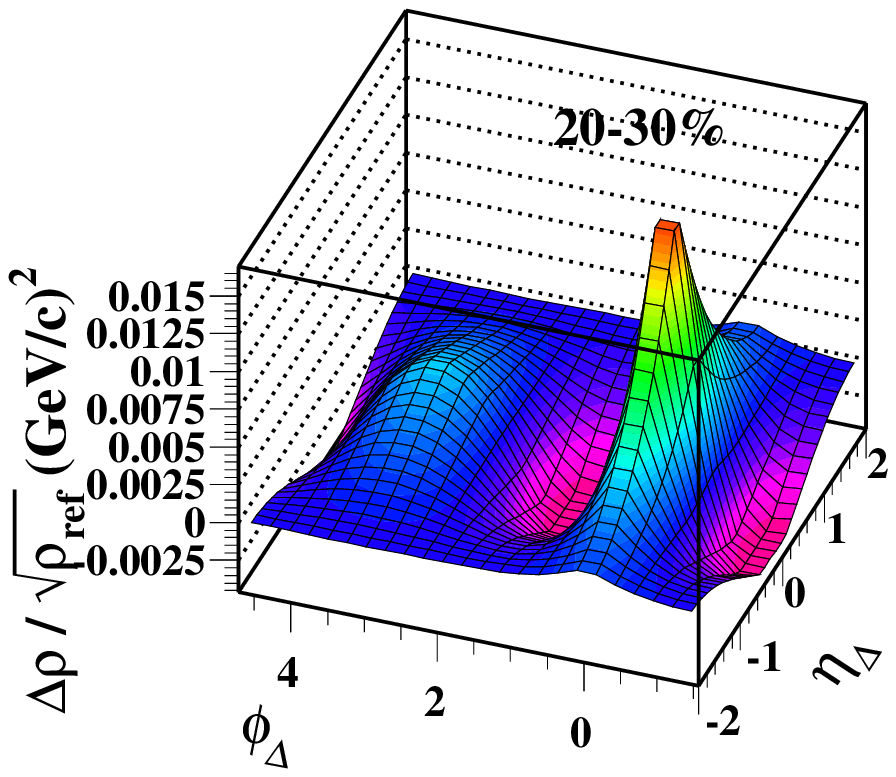}
\end{minipage} & 
\begin{minipage}{.47\linewidth}
\includegraphics[keepaspectratio,width=1.65in]{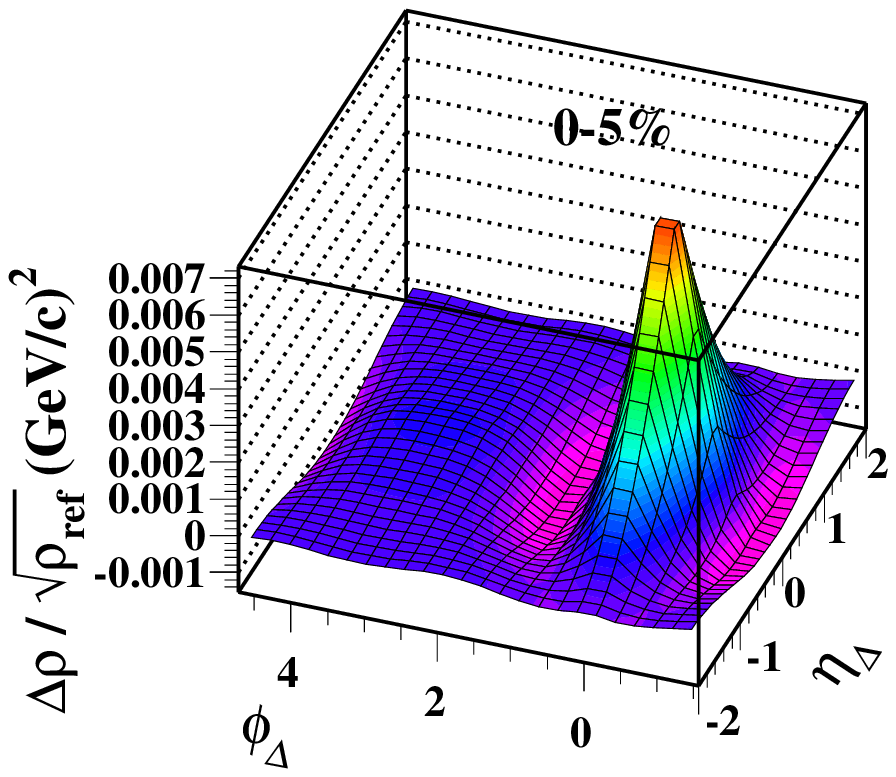}
\end{minipage}\\
\end{tabular}
\caption{Distributions of $\Delta \rho / \sqrt{\rho_{ref}}(\eta_\Delta,\phi_\Delta)$  for 80-90\% (upper left), 45-55\% (upper right),  20-30\% (lower left) and 0-5\% (lower right) of total cross section. Monopole (constant), dipole and quadrupole components have been subtracted.   \label{fig2}}
\end{figure}

In Fig.~\ref{fig2} monopole (constant offset), dipole $\cos(\phi_\Delta)$ and quadrupole $\cos(2\phi_\Delta)$ components (sinusoids evident in Fig.~\ref{fig1}, right panels) have been subtracted from the autocorrelations for centralities (80-90, 45-55, 20-30 and 0-5\%) by minimizing residuals of the three components on the away side ($|\phi_\Delta| > \pi / 2$) and for $|\eta_\Delta| \sim 2$ (minimizes influence of away-side peak structure). The full-$\phi$-acceptance fluctuations in Fig.~\ref{fig1} (left panels) are determined only by the minijet structure and the monopole component; the dipole and quadrupole components integrate to zero at $\delta \phi = 2\pi$. Since one interpretation of $p_t$ correlations is that they reflect velocity correlations of local particle source velocities, the quadrupole component from this analysis may constitute the first observation of elliptic flow as a {\em true velocity phenomenon}. 

The subtracted autocorrelations retain three structures localized on both $\eta_\Delta$ and $\phi_\Delta$: a same-side positive peak, a same-side negative peak (apparent as the regions of {\em negative} correlation immediately adjacent to the positive peak on $\phi_\Delta$) and an away-side peak. The near-side negative peak cannot be a result of incorrect subtraction of the multipole components. The latter have by definition no structure (are constant) on $\eta_\Delta$, whereas the negative near-side peak is highly structured (a peak rising symmetrically to zero) on $\eta_\Delta$. The near-side positive peak, in the absence of alternative explanations, is interpretable as a velocity correlation associated with semi-hard parton scattering (minijets). Those three $p_t$ correlation structures comprise the main subject of this paper.
 

In Fig.~\ref{fig2} we observe that the three peak features vary strongly in shape and amplitude with collision centrality. For the more central collisions we observe that the same-side positive peak is substantially elongated along $\eta_\Delta$ and significantly {\em narrowed} along $\phi_\Delta$. We quantify those observations with model fits. The autocorrelations in Fig.~\ref{fig2} were fitted with the model function defined in Eq.~(\ref{fit}), a sum of near-side positive peak $B_1$, near-side negative peak $B_2$ (signed number) and away-side peak $B_3$ 
, each with the same form,
\bea \label{fit}
F &=&  \sum_{i=1}^3B_i\, \exp\left\{-\left|\frac{\eta_\Delta}{\sqrt{2}\, \sigma_{\eta i}}\right|^{\tau_{\eta i}} \hspace{-.15in} -\left|\frac{\phi_\Delta  - \delta_{i3}\,\pi}{\sqrt{2}\, \sigma_{\phi i}}\right|^{\tau_{\phi i}}\right\}, 
\eea
where $\delta_{i3}$ is a Kronecker delta. 
 This function includes exponents $\tau$ as shape parameters. In contrast to a gaussian ($\tau \equiv 2$), which best describes near-side peaks for number autocorrelations~\cite{axialci}, best-fit exponents for these $p_t$ autocorrelations were found to be $\tau_{\eta 1} = \tau_{\eta 3} = \tau_{\phi 1} =1.5\pm0.1$, with $\tau_{\phi 2} = 2.5\pm0.1$, $\tau_{\phi 3} = 1.9\pm0.1$ and $\tau_{\eta 2}  =1.7\pm0.1$ (for all centralities). Widths for near-side negative and away-side peaks varied (respectively from peripheral to central) nearly linearly over the ranges $0.75 < \sigma_{\eta 2} < 1.1\pm0.1$, $ 0.9 < \sigma_{\eta 3} < 1\pm0.1$,  $\sigma_{\phi 2} \sim 2.1\pm0.2$ and $2.4 > \sigma_{\phi 3} > 1.5\pm0.1$. 

\begin{figure}[h]
\includegraphics[keepaspectratio,width=3.3in]{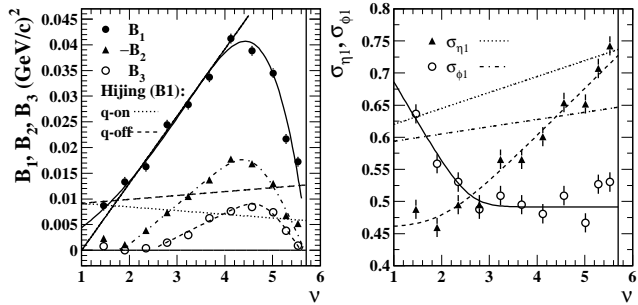}
\caption{Correlation amplitudes (left panel) plotted {\em vs} mean participant path length $\nu$. Solid dots are for positive near-side peaks, solid triangles are for negative near-side peaks. Open circles are for away-side peaks. Positive-peak widths (right panel) are plotted on $\nu$ for pseudorapidity (solid triangles) and azimuth (open circles). Curves guide the eye (see text). Dotted, dashed and dash-dot lines represent Hijing results. \label{fig3}}
\end{figure}

The best-fit amplitudes for all peaks, corrected for background contamination and tracking inefficiency (21-38\% for these 200 GeV data, increasing from minimum to maximum centrality)~\cite{ptprl} and widths for the near-side positive peak are plotted in Fig.~\ref{fig3} {\em vs} mean participant path length $\nu$~\cite{nu,central}. The vertical line to the right of each panel indicates the estimated limit of $\nu$ for Au-Au collisions, corresponding to $b \sim 0$ and $N_{part} / 2 \sim 191$. The fitted peaks are strongly localized on $\eta_\Delta$, and very different in shape from the $\eta_\Delta$-independent sinusoid components subtracted from autocorrelations in Fig.~\ref{fig1} (right panels) to form those in Fig.~\ref{fig2}. There is therefore negligible cross-talk between the two types of fitting function. The difference between fitting the data in Fig.~\ref{fig1} (right panels) including sinusoids in the model function and the data in Fig.~\ref{fig2} with Eq.~(\ref{fit}) is less than the stated errors in the fit parameters. Residuals from fitted peak structures were typically less than 2\% of the near-side peak amplitude. The peak amplitudes increase with centrality to a maximum value and then decrease sharply for the most central collisions. The near-side positive-peak width on $\eta_\Delta$ increases monotonically with centrality, while that on $\phi_\Delta$ decreases. 
The rising part of the $B_1$ data is consistent with both a straight line $\propto (\nu - 1)$ and a curve $\propto \nu^{1.6}$. 


The autocorrelation density ratio $\Delta \rho(p_t:n) / \sqrt{\rho_{ref}(n)}$ measures {\em relative covariances} (proportional to Pearson's correlation coefficient) of $\langle p_t \rangle $ fluctuations at pairs of points on $(\eta,\phi)$ separated by $(\eta_\Delta,\phi_\Delta)$. The autocorrelation distribution reveals the {\em average} shape of localized structures which may appear in different places on $(\eta,\phi)$ in different events, and possibly in only some events, but which have some shape stability over an event ensemble. We observe substantial (0.01 - 0.1) $p_t$ covariances which can be interpreted as local transverse-velocity and/or temperature fluctuations correlated at pairs of points on $(\eta,\phi)$. As with the separation of temperature and transverse-flow aspects of inclusive single-particle $p_t$ distributions, temperature and velocity correlations may also be distinguished as to source mechanism if mass identification is implemented in $p_t$ fluctuation/correlation analysis. 

The structures in Figs.~\ref{fig1} (right panel) and \ref{fig2} can be compared to signature angular correlations for high-$p_t$ parton fragments: a same-side 2D peak at the origin (jet cone) and an away-side $\eta_\Delta$-invariant ridge on $\phi_\Delta$ (apparent as such only in 2D analyses like this one). The correlations in Fig.~\ref{fig1} (upper-right panel) have exactly that structure, and agree in detail with the Hijing correlations described below which are known to represent low-$p_t$ jets, providing strong evidence that the dominant source of $p_t$ correlations for p-p and peripheral Au-Au collisions is low-$p_t$ parton fragments.  For low-$p_t$ partons the away-side ridge in Fig.~\ref{fig1} (upper-right panel) is not distinguishable from one lobe of a sinusoid, and is thus removed in the multipole subtraction to obtain Fig.~\ref{fig2}. The same-side positive peak then represents the conventional jet-cone structure, albeit in $p_t$ rather than angular correlations. We can then argue that the same features for more central Au-Au collisions continue to derive from low-$p_t$ partons, but with modifications by a colored dissipative medium.

The model-fit results in Fig.~\ref{fig3} illustrate the dramatic changes in Fig.~\ref{fig2} structure with collision centrality. Focusing on the near-side positive peak, the width on $\phi_\Delta$ falls by 30\%, whereas the width on $\eta_\Delta$ increases by 60\%. Those trends are qualitatively similar to equivalent measurements of angular correlations~\cite{axialci}, where the large width increase on $\eta_\Delta$ was interpreted as due to strong coupling of low-$p_t$ partons to the longitudinally-expanding colored medium. The $\sigma_\eta$ variation is much less for $p_t$ correlations, suggesting that elongation of parton fragment distributions on $\eta$ involves lower-$p_t$ particles with increasing $\eta_\Delta$. The near-side peak for $p_t$ correlations is significantly non-gaussian , the sharper peak represented by exponent $\tau = 1.5$ (the correspondent for angular correlations is a gaussian with $\tau = 2$). 

The amplitude $B_1$ of the near-side positive peak increases by a factor four or more with centrality, but falls off rapidly for the most central collisions, in contrast to the subtracted monopole term which increases monotonically to mid centrality and then remains constant with further centrality increase. Since the relative covariance could also be interpreted (with strong assumptions) as a number of correlated pairs per particle in the system, the increase of $B_1$ may indicate that the number of correlated pairs from minijets increases {\em relative to the total multiplicity}. If the system were composed only of {\em independent}  minijets (with no soft particle production) the autocorrelation density ratio (and variance difference) would be independent of system size (minijet number). The observed increase with system size could then result from a larger number of low-$p_t$ partons, a larger mean fragment multiplicity for each parton, or both. The other correlation structures, the negative same-side and positive away-side peaks, are unique to $p_t$ correlations and will be considered in detail in a followup publication. The presence of the negative near-side peak means that the variance difference, as an integral fluctuation measure, significantly underestimates the relative amount of minijet structure. 





An equivalent analysis of $\langle p_t \rangle$ fluctuations in Hijing collisions~\cite{qingjun} indicates that the near-side peak shapes for Hijing (quench on or quench off) are nearly symmetric on $(\eta_\Delta,\phi_\Delta)$, with shape described by single exponent $\tau = 1.7\pm0.1$. The combination of same-side 2D peak and away-side 1D azimuthal ridge observed in that analysis supports the interpretation that the basic source of those $p_t$ correlations is low-$p_t$ parton fragments or minijets, consistent with the basic pQCD jet model in Hijing. The centrality dependence of the amplitudes of the Hijing near-side peak for quench-on (default) and quench-off collisions is represented respectively by dotted and dashed lines in Fig.~\ref{fig3} (left panel). The lines in the right panel correspond to Hijing default (quench on) same-side peak widths. The amplitude (width) centrality trends for default Hijing are similar: modest variations linear with path length $\nu$. The differences between quench-on and quench-off results from central Au-Au collisions for amplitudes and widths, representing pQCD modeling of in-medium parton energy loss, are $\sim 10$\%. 

Comparing Hijing to the present analysis in Fig.~\ref{fig3} we note that there are at least three instances of qualitative disagreement between Hijing and RHIC data. First, for very peripheral Au-Au collisions (and therefore nucleon-nucleon collisions) the Hijing same-side peak (jet cone) is symmetric on angle, whereas the data are quite asymmetric. Thus, there is already disagreement with data at the level of parton fragmentation in elementary collisions. Second, the centrality dependencies of the amplitudes and widths of the same-side peak are qualitatively different from data: in some cases even the sign of the variation is wrong. Third, the amplitude of the same-side Hijing peak is qualitatively less than that for data in mid-central collisions. The last is especially surprising when comparing quench-off Hijing (dashed line in the left panel) to data. The quench-off Hijing option in principle models jet production from a linear superposition of N-N collisions combined with a Glauber model of a {\em transparent} nucleus. That model should provide an upper limit for jet structure in A-A collisions. Yet the same-side peak amplitude for quench-off Hijing is 2-3 times {\em less} than that for mid-central RHIC Au-Au collisions which are observed to be {\em highly opaque} to minijets in the central region.  Finally, we observe no evidence in Hijing data for the same-side negative peak which is a prominent new feature of RHIC data. The perturbative treatment of parton energy loss in Hijing appears to disagree strongly with the observed process for the minimum-bias partons which dominate $p_t$ correlations in Au-Au collisions.



In conclusion, we have for the first time measured the scale dependence of $\langle p_t \rangle$ fluctuations on $(\delta \eta,\delta \phi)$ in heavy ion collisions. We have inverted those distributions to obtain autocorrelation distributions on corresponding difference variables $(\eta_\Delta,\phi_\Delta)$ which reveal the correlation structure of the local properties of a two-particle $p_t$ distribution, specifically a combination of local transverse velocity and temperature. Inferred autocorrelations reveal complex $p_t$ correlation structure in Au-Au collisions at RHIC, including peaked structures attributed to minijets which vary strongly with collision centrality. We observe that $p_t$ autocorrelations provide unique access to minijet structure down to very low $p_t$ and probe the detailed interplay between low-$p_t$ partons and the dissipative colored medium. Further studies with identified particles may separately characterize the local velocity and temperature structures of heavy ion collisions.


We thank the RHIC Operations Group and RCF at BNL, and the
NERSC Center at LBNL for their support. This work was supported
in part by the HENP Divisions of the Office of Science of the U.S.
DOE; the U.S. NSF; the BMBF of Germany; IN2P3, RA, RPL, and
EMN of France; EPSRC of the United Kingdom; FAPESP of Brazil;
the Russian Ministry of Science and Technology; the Ministry of
Education and the NNSFC of China; IRP and GA of the Czech Republic,
FOM of the Netherlands, DAE, DST, and CSIR of the Government
of India; Swiss NSF; the Polish State Committee for Scientific 
Research; STAA of Slovakia, and the Korea Sci. \& Eng. Foundation.



\begin{thebibliography}{9}

\bibitem{QCD} J.~C.~Collins and M.\,J.~Perry, Phys. Rev. Lett. {\bf 34}, 1353 (1975);
	B.~Freedman and L.~McLerran, Phys. Rev. {\bf D 17}, 1109 (1978);
	G.~Baym and S.~A.~Chin, Phys. Lett. B {\bf 62}, 241 (1976);
N.~Cabibbo and G.~Parisi, Phys. Lett. B {\bf 59} (1975) 67;
Proceedings of Quark Matter 2004, J. Phys. G {\bf 30}, S633 (2004).

\bibitem{theor0}
K.~Kajantie, P.~V.~Landshoff and J.~Lindfors,
Phys.\ Rev.\ Lett.\  {\bf 59}, 2527 (1987).

\bibitem{theor1} 
A.~H.~Mueller,
Nucl.\ Phys.\ B {\bf 572}, 227 (2000).

\bibitem{theor2} 
G.~C.~Nayak, A.~Dumitru, L.~McLerran and W.~Greiner,
Nucl.\ Phys.\ A {\bf 687}, 457 (2001).

\bibitem{qgp} D. d'Enterria, nucl-ex/0309015.
 
\bibitem{backjet} C.~Adler {\em et al.} (STAR Collaboration), Phys. Rev. Lett. {\bf 90}, 082302 (2003).

\bibitem{suppress}
C.~Adler {\it et al.} (STAR Collaboration),
Phys.\ Rev.\ Lett.\  {\bf 89}, 202301 (2002).

\bibitem{suppress2}
J.~Adams {\it et al.}  (STAR Collaboration),
Phys.\ Rev.\ Lett.\  {\bf 91}, 172302 (2003).

\bibitem{Phenix} K. Adcox et al., Phys. Rev. {\bf C 66}, 024901 (2002).

\bibitem{ptprl} J. Adams {\it et al.} (STAR Collaboration), Phys. Rev. C {\bf 71}, 064906 (2005). 

\bibitem{QT} Q.\,J.~Liu and T.\,A.~Trainor, Phys. Lett. B {\bf 567}, 184 (2003). 


\bibitem{inverse} T.\,A.~Trainor, R.\,J.~Porter and D.\,J.~Prindle, J. Phys. G: Nucl. Part. Phys.  {\bf 31} 809 (2005). 

\bibitem{mtxmt} J. Adams {\it et al.} (STAR Collaboration), nucl-ex/0408012.  

\bibitem{cltps} T.\,A.~Trainor, eprint hep-ph/0001148.

\bibitem{axialcd} J. Adams {\it et al.} (STAR Collaboration), Phys. Lett. B {\bf 634}, 347 (2006). 

\bibitem{axialci} J. Adams {\it et al.} (STAR Collaboration), nucl-ex/0411003. 
 
\bibitem{qingjun} Q.\,J.~Liu, D.\,J.~Prindle and T.\,A.~Trainor, Phys. Lett. B {\bf 632}, 197 (2006). 

\bibitem{highptphi} $p_t$ correlation measure $\Delta \rho(p_t:n) / \sqrt{\rho_{ref}(n)}$ is related to the number-correlation measure in~\cite{axialci,axialcd} by $\bar N\, (\hat r - 1) \simeq 2\Delta \eta \Delta \phi \sqrt{\rho_{ref}}\, \Delta \rho / \rho_{ref} \simeq 24 \Delta \rho(n) / \sqrt{\rho_{ref}(n)}$. It is also related to single-particle {\em conditional} distribution $1 / N_{trig}\, dN/d\Delta \phi$~\cite{backjet} but invokes no trigger condition.

\bibitem{pearson} J.\,F.~Kenney and E.\,S.~Keeping, Mathematics of Statistics, Pt. 1, 3rd ed.  Princeton, NJ: Van Nostrand, 1962.

\bibitem{mitinv} T.\,A.~Trainor and D.\,J.~Prindle (STAR Collaboration),  Proceedings of the MIT Workshop on Correlations and Fluctuations in Relativistic Nuclear Collisions, Cambridge, Massachusetts, 21-23 April 2005, J. Phys.: Conference series, {\bf 27}, 118 (2005).

\bibitem{starnim}
K. H. Ackermann {\em et al.}, Nucl. Instrum. Meth. A {\bf 499}, 624 (2003);
see other STAR papers in volume A{\bf 499}.


\bibitem{spectra} C. Adler {\em et al.} (STAR Collaboration), Phys. Rev. Lett. {\bf 87}, 112303
(2001); {\em ibid}. {\bf 89}, 202301 (2002).


\bibitem{central} T.\,A.~Trainor and D.\, J.~Prindle, hep-ph/0411217. 

\bibitem{nu} $\nu$ 
estimates the mean participant path length in number of encountered nucleons~\cite{central}.














\end{thebibliography}
\end{document}